\documentclass[11pt]{article}
\usepackage[margin=1in]{geometry}
\usepackage{authblk}
\usepackage{setspace}
\usepackage{mathtools}
\usepackage{cite}
\usepackage{url}
\usepackage{graphicx}
\usepackage{amsmath}
\usepackage{amssymb}
\usepackage{braket}
\usepackage{bm}
\usepackage{booktabs}
\usepackage{xcolor}

\usepackage{caption}
\DeclareCaptionLabelSeparator{bar}{ \textbar\ }
\usepackage[normalem]{ulem}

\begin{document}

\title{Programmable nonlinear function synthesis on a photonic processor with quantum signal processing}
\author[1]{Elaheh Karooby\thanks{These authors contributed equally.}}
\author[1]{Masoud Hakimi Heris\protect\footnotemark[1]}
\author[1,2,3]{Yuan Liu\thanks{Corresponding author: q\_yuanliu@ncsu.edu}}
\author[1,2]{Qing Gu\thanks{Corresponding author: qgu3@ncsu.edu}}

\affil[1]{Department of Electrical and Computer Engineering, North Carolina State University, Raleigh, NC, USA}
\affil[2]{Department of Physics and Astronomy, North Carolina State University, Raleigh, NC, USA}
\affil[3]{Department of Computer Science, North Carolina State University, Raleigh, NC, USA}

\date{}


\maketitle
\begin{abstract}
Programmable photonic integrated circuits are emerging as increasingly large and versatile interferometric processors operating at room temperature. However, their native operations are linear, while many computational tasks require nonlinear input–output transformations that typically rely on nonlinear optical materials or resonant devices. Here, we establish a direct mapping between quantum signal processing (QSP) and the native two-mode operations of programmable interferometric photonic integrated circuits, showing that the $SU(2)$ structure required by QSP can be realized through programmable phase control and mode mixing. We experimentally demonstrate this correspondence using dual-rail single-photon encoding on a 24-mode programmable photonic integrated circuit, realizing QSP sequences up to depth $L=11$ and synthesizing STEP, ReLU and SELU functions through programmable phase control. The optical transformation remains linear, while repeated encoding of the input variable and coherent interference produce a nonlinear dependence of the output probabilities on the encoded variable. Across all accessible circuit depths, the measured responses follow the programmed QSP transformations, with hardware-induced mean squared errors between $10^{-3}$ and $10^{-2}$, all remaining below the intrinsic finite-depth approximation error. These results establish an algorithm-to-hardware mapping of QSP on programmable photonic processors. They also demonstrate an algorithmic route to programmable nonlinear functions of encoded variables on linear photonic hardware.
\end{abstract}

\section{Introduction}
Programmable photonic integrated circuits (PICs) are emerging as versatile platforms for quantum information processing, combining room-temperature photonic processing with compact integration and reconfigurable control of optical transformations~\cite{Wang2020IQP,Pelucchi2022,Aharonovich2026,Clements2016}. Increasingly large programmable photonic processors can implement complex unitary transformations through reconfigurable networks of phase shifters and beam splitters, providing a reconfigurable platform for information processing~\cite{Bogaerts2020,Carolan2015,Taballione2023,Barzaghi2025,Chi2022,Maring2024}. However, scaling photonic hardware alone does not establish how these native operations can be used to realize more complex computational transformations. This limitation becomes especially relevant for nonlinear input-output mappings. Programmable linear optics naturally implements unitary transformations, whereas nonlinear functionality typically relies on nonlinear optical materials, resonant optical devices, optoelectronic conversion, or measurement and electronic feedback~\cite{Williamson2020,Shen2017}. An alternative is to encode such functionality at the algorithmic level, using a computational framework that maps directly onto the native operations of programmable photonic hardware. Such a framework could extend the functionality of linear-optical processors beyond their native unitary transformations without requiring physical optical nonlinearities, while providing a much-needed, scalable quantum architecture for photonic quantum processors that could bridge the gap between hardware and utility-scale computational workloads.

Quantum signal processing (QSP) provides such a theoretical framework. QSP synthesizes polynomial transformations of an encoded variable~\cite{low2016,low2017} through sequences of programmable single-qubit phase rotations. QSP forms the foundation of quantum singular value transformation (QSVT), which enables a broad class of quantum algorithms, including Hamiltonian simulation, quantum linear-system solvers, amplitude amplification and Gibbs-state preparation~\cite{low2019,Gilyen2019,Martyn2021,dong2022,Rossi2023,joven2026scalable}. QSP has also been explored in quantum machine learning~\cite{Biamonte2017,bu2025,Guo2024} and quantum sensing~\cite{sinanan2024single,majumdar2026robust}, where polynomial transformations provide expressive nonlinear mappings and data processing. Despite its broad algorithmic relevance, experimental demonstrations of QSP-based protocols have so far been restricted primarily to matter-qubit platforms, including QSVT-based benchmarking on cryogenic superconducting processors~\cite{dong2022} and trapped-ion systems where high-fidelity quantum control enables accurate implementation of phase sequences~\cite{bu2025,Kikuchi2023}. As circuit depth increases, these implementations face operational errors and decoherence~\cite{bu2025}, motivating alternative hardware platforms capable of implementing increasingly complex QSP protocols while supporting room-temperature photonic processing.

Programmable interferometric PICs are particularly well matched to QSP because the elementary operations required by the algorithm, namely phase rotations and mode mixing, map directly onto optical phase shifters and Mach–Zehnder interferometers. This correspondence allows a desired function to be synthesized into a sequence of phase settings and executed directly on a programmable linear-optical mesh. Although the underlying PIC remains strictly linear, repeated encoding of the input variable, coherent interference, and quantum measurement together yield nonlinear input-output mappings whose functional form is determined algorithmically by the programmed phase sequence. QSP therefore provides a route to programmable nonlinear functions of an encoded input variable without requiring nonlinear optical materials or dedicated nonlinear activation devices. This direct algorithm-to-hardware mapping suggests a broader role for QSP as an algorithmic framework and architectural backbone for photonic information processing.

Here, we establish a direct mapping between the single-qubit operations of QSP and the native two-mode transformations of programmable interferometric PICs, and experimentally realize this mapping on a programmable linear photonic processor. Using a 24-mode programmable processor and dual-rail single-photon encoding, we implement QSP sequences with circuit depths up to $L=11$ and experimentally synthesize representative nonlinear functions including STEP, ReLU and SELU. The measured responses closely follow the analytic QSP predictions and hardware-level simulations across all experimentally accessible QSP sequence lengths supported by the photonic processor. We further analyze the distinct error mechanisms governing photonic and matter-qubit QSP implementations. In the photonic implementation, photon loss is heralded and therefore primarily reduces the successful sampling rate, while residual phase-shifter errors can be mitigated through classical calibration. Consistent with this error structure, the measured hardware-induced error remains below the intrinsic finite-depth approximation error and shows no systematic increase with circuit depth over the experimentally accessible range. These results establish a direct mapping between QSP and programmable interferometric photonics and demonstrate an algorithmic approach for generating programmable nonlinear input–output mappings on linear photonic hardware.

\section{Photonic implementation of QSP}
\label{sec:implementation}

\begin{figure}[!t]
  \centering
  \includegraphics[width=\textwidth]{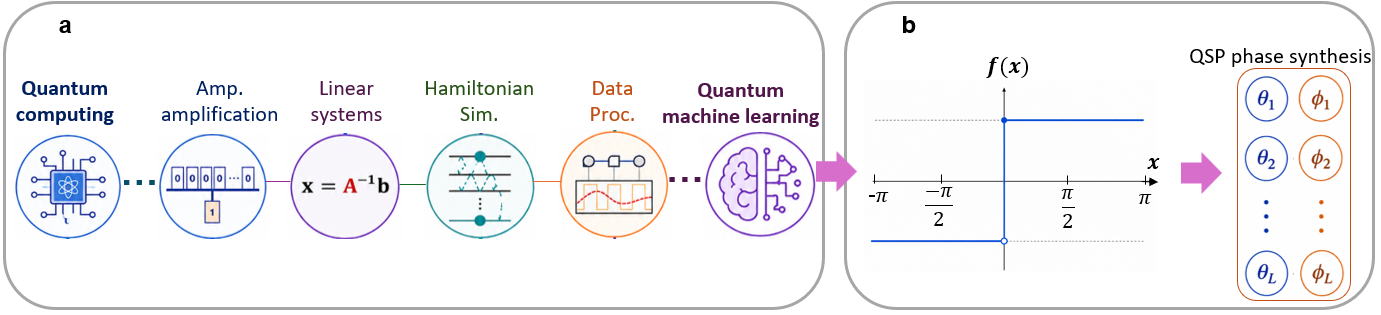}%
  \caption{\label{fig:architecture}
\textbf{QSP applications and target nonlinear function synthesis.}
\textbf{a,} Representative application areas of QSP, ranging from quantum
algorithms to quantum machine learning.
\textbf{b,} A target nonlinear function is approximated by a finite
trigonometric polynomial, and the corresponding QSP phase sequence
$\{(\theta_j,\phi_j)\}_{j=0}^{L}$ is computed classically.}
\end{figure}

\begin{figure}[!t]
  \centering
  \includegraphics[width=\textwidth]{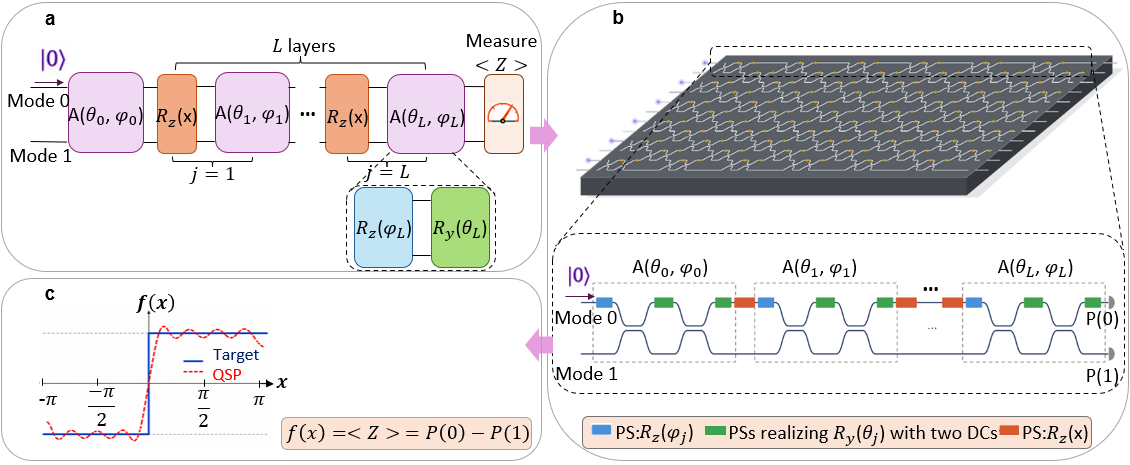}%
  \caption{\label{fig:qsp_PIC}
  \textbf{Algorithm-to-hardware mapping of QSP on a programmable PIC.}
  \textbf{a,} Trigonometric QSP circuit. 
  Eq.~(\ref{eq:qsp}): an initial signal-processing unitary
  $A(\theta_0,\phi_0)$ followed by $L$ layers, each interleaving the
  signal unitary $R_z(x)$ with a signal-processing unitary
  $A(\theta_j,\phi_j)$. The output is measured through
  $\langle Z\rangle=P(0)-P(1)$.
  \textbf{b,} Direct mapping of the QSP sequence onto native two-mode operations of a programmable interferometric PIC.
  Logical states are path-encoded, with the upper and lower waveguides
  representing $|0\rangle$ and $|1\rangle$, respectively.
  Each $A(\theta_j,\phi_j)$ is implemented by an MZI comprising two
  nominally 50:50 directional couplers (DCs) and programmable phase
  shifters (PSs), while $R_z(x)$ is implemented through the phase
  difference between the two waveguides. The circuit is programmed on a
  Clements-style rectangular mesh~\cite{Clements2016}.
  \textbf{c,} QSP approximation of the target nonlinear function. The finite-depth
  QSP sequence produces the characteristic trigonometric approximation,
  and the synthesized function is reconstructed from the measured output
  probabilities as $f(x)=\langle Z\rangle=P(0)-P(1)$.}
\end{figure}

QSP implements polynomial transformations through alternating signal
encoding and programmable single-qubit rotations~\cite{low2016}.
In this work, we first establish that these operations can be mapped directly onto the native two-mode transformations of a programmable interferometric PIC. Specifically, optical phase differences implement the required $R_z$ rotations, while programmable MZIs realize the required $R_y$ rotations. The complete $SU(2)$ correspondence is derived in Methods and Supplementary Section 1. We then adopt the trigonometric formulation of
QSP~\cite{wang2023qpp}, because its signal variable is encoded as a phase rotation that maps naturally onto the programmable phase difference in an interferometric PIC.
For a QSP circuit of
depth $L$, the QSP unitary is given by
\begin{equation}
W(x)
=
A(\theta_0,\phi_0)
\prod_{j=1}^{L}
R_z(x)A(\theta_j,\phi_j),
\label{eq:qsp}
\end{equation}
where $A(\theta,\phi)=R_y(\theta)R_z(\phi)$ is the signal-processing unitary, $R_y(\theta)$ and $R_z(\phi)$ denote single-qubit rotations about the $y$- and $z$-axes with angles $\theta$ and $\phi$, respectively, and the input variable $x$ is repeatedly encoded through the signal unitary $R_z(x)$ at each layer. Here, $x$ denotes a dimensionless scalar input physically encoded as the phase difference between the two optical modes, rather than an optical field amplitude or intensity. In applications, an external variable can be mapped to this phase-encoding domain. The rotation angles $\{(\theta_j,\phi_j)\}_{j=0}^{L}$ are determined classically to approximate the desired target function. The approximated function is given by the expectation value of the Pauli-$Z$ operator, $f(x) \coloneqq \langle Z\rangle = P(0)-P(1)$, where $P(0)$ and $P(1)$ denote the probabilities of measuring the two computational basis states. Figure~\ref{fig:architecture} shows the target of the QSP procedure,
including representative applications and the classical synthesis of a
desired nonlinear function into the QSP rotation-angle sequence
$\{(\theta_j,\phi_j)\}_{j=0}^{L}$.
Figure~\ref{fig:qsp_PIC} shows how this sequence is realized experimentally,
from the trigonometric QSP circuit and its implementation on the
programmable PIC to reconstruction of the output function from the measured
probabilities.

Unlike trapped-ion implementations, where the QSP sequence is
realized through laser and microwave control pulses~\cite{bu2025}, the
operations entering Eq.~(\ref{eq:qsp}) map directly onto two-mode
interferometric transformations, as shown in Methods. Logical qubit states are path-encoded, with the upper and lower waveguides respectively representing the logical basis states $|0\rangle$ and $|1\rangle$ [Fig.~\ref{fig:qsp_PIC}(b)]. The complete circuit is programmed on a Clements-style rectangular mesh of MZIs~\cite{Reck1994,Clements2016,Carolan2015}, which supports arbitrary linear transformations by adjusting the phase shifters.

The experiments were performed on the 24-mode Belenos programmable
photonic processor, accessed through Quandela's cloud platform~\cite{Maring2024}
(see Methods for platform details). We program our trigonometric QSP circuit on the two top waveguide channels. For each input $x\in[-\pi,\pi]$, the phase shifters that implement the signal unitary $R_z(x)$ are programmed to encode the input into the PIC. A single photon is then injected into the upper waveguide of the dual-rail circuit, and the output probabilities $P(0)$ and $P(1)$ are extracted from repeated single-photon measurements at the two output ports. The output is reconstructed as $f(x)=P(0)-P(1)$. By sweeping $x$ across the input domain, the complete QSP function approximation is obtained.

\section{Nonlinear function synthesis on a linear photonic processor}
We experimentally synthesize three nonlinear functions on the
photonic processor, namely STEP, ReLU, and SELU. These functions
are of broad relevance in quantum algorithms and machine learning: the
STEP function underlies eigenphase classification in quantum phase
estimation~\cite{bu2025,wang2023qpp}, while ReLU and SELU are standard neural-network activation
functions for universal function
approximation~\cite{nair2010,glorot2011,klambauer2017}. Following the trigonometric QSP convention in
Sec.~\ref{sec:implementation}, each target function is defined
on the interval $x\in[-\pi,\pi]$ and approximated by a degree-$L$ trigonometric
polynomial realized by Eq.~(\ref{eq:qsp}). The classical computation of the rotation angles
$\{(\theta_j,\phi_j)\}_{j=0}^{L}$ is described in the Methods, with
target-function definitions and phase sequences provided in
the Supplementary Information.

Figure~\ref{fig:functions} shows the experimental realization of the three
functions. In the rectangular mesh of the 24-mode processor, 12 MZIs act on the programmed pair of modes in alternating columns; since a depth-$L$ QSP circuit requires $L{+}1$ signal-processing unitaries, this constrains our implementation to a maximum circuit depth of $L=11$. We therefore present results at $L=3$, $7$, and $11$ to illustrate the
effect of increasing the circuit depth. For each input value $x$, the
reconstructed output $f(x)=P(0)-P(1)$ is obtained from repeated
heralded single-photon measurements, and $x$ is swept across the input
domain of $[-\pi,\pi]$ to trace the complete approximation. In addition to the experimentally obtained output function,
we plot the mathematical target function, the analytic prediction of the
QSP, and the numerically simulated PIC performance. For STEP, the smooth surrogate used for angle finding is also
shown. For all three functions, the measured data closely follow the
analytic QSP prediction. Because a QSP sequence of depth $L$ realizes a
trigonometric polynomial supporting harmonics up to order $L$, increasing
the depth from $L=3$ to $L=11$ enriches the accessible Fourier spectrum and improves the approximation accuracy, with the measured curves converging
toward the target function as additional layers are included.

The nonlinear functions in Figure~\ref{fig:functions} do not arise from an optical nonlinearity. The photonic processor itself remains strictly linear, with its beam splitters and phase shifters performing only linear transformations of the optical modes. Instead, the nonlinear dependence on the encoded variable is synthesized by the QSP sequence. Repeated encoding of the same input $x$ across $L$ layers generates a degree-$L$ trigonometric polynomial in $x$~\cite{gilvidal2020}, while measurement converts the resulting output amplitudes into the probabilities used to evaluate $f(x)=P(0)-P(1)$. The resulting nonlinear dependence of $f(x)$ on the encoded variable therefore arises from the QSP sequence and readout rather than from a nonlinear optical material or device. This differs from conventional photonic neural-network implementations, where nonlinear activation is typically introduced through optoelectronic processing or material nonlinearities~\cite{Shen2017,Williamson2020}. Our results show that nonlinear function approximations can be synthesized algorithmically using the native operations of a programmable linear-optical mesh. 

\begin{figure}[!t]
  \centering
  \includegraphics[width=\textwidth]{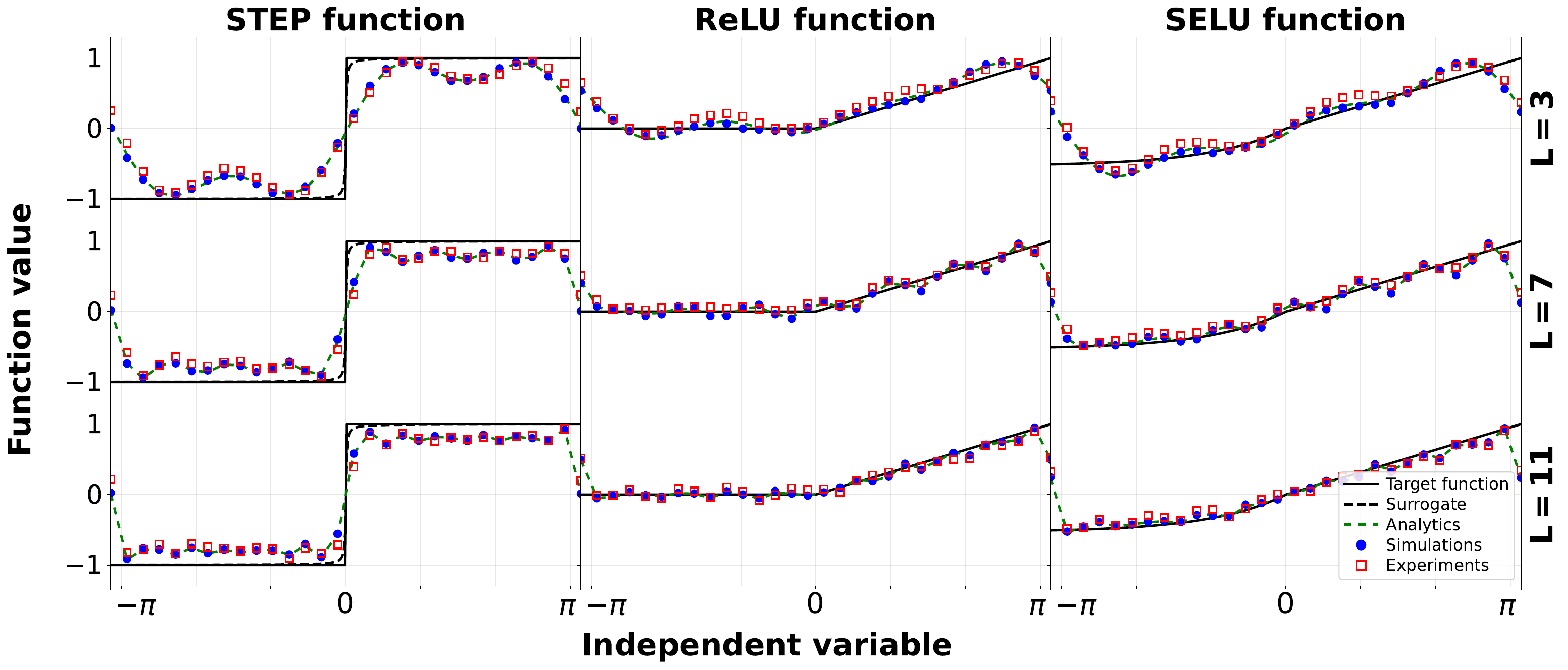}
  \caption{\label{fig:functions}\textbf{Experimental realization of STEP,
ReLU, and SELU functions by photonic QSP.}
Reconstructed function $f(x)=P(0)-P(1)$ versus input $x$ for
STEP (left column), ReLU (middle column), and SELU (right column) target
functions, each shown at QSP circuit depths of $L=3$ (top row), $L=7$
(middle row), and $L=11$ (bottom row). In each panel, the solid black
line is the ideal target function, the dashed green line is the analytic prediction of the QSP, blue circles represent
numerically simulated PIC behavior, and red squares show
experimental results. Experimental data points are reconstructed
from $N_{\mathrm{shots}}=5000$ heralded single-photon detection events
per input value. For all three functions, the measured data closely follow the
analytic prediction of QSP, and the approximation accuracy of the target function
improves with increasing depth $L$.}
 \vspace{-4mm}
\end{figure}

\section{Error analysis and platform comparison}
\label{sec:erroranalysis}
To quantify the accuracy of each realization, we compute the mean-squared
error (MSE) between the approximated function at depth $L$ and the target
function,
\begin{equation}
\mathrm{MSE}_L=\frac{1}{N}\sum_{n=1}^{N}
\bigl|\,f_L(x_n)-f_{\mathrm{target}}(x_n)\,\bigr|^{2},
\label{eq:mse}
\end{equation}
where $f_L(x_n)$ is the reconstructed output at input $x_n$ for a QSP circuit
of depth $L$, obtained analytically from Eq.~(\ref{eq:qsp}), from the
numerical PIC simulation, or from the experiment, and
$f_{\mathrm{target}}$ is the target function, namely the smooth surrogate for STEP
and the original functions for ReLU and SELU. Averaging the squared deviations
over $N$ input values sampled uniformly across $x\in[-\pi,\pi]$ yields the
MSE at each depth $L$.

\begin{figure}[!t]
  \centering
  \includegraphics[width=\textwidth]{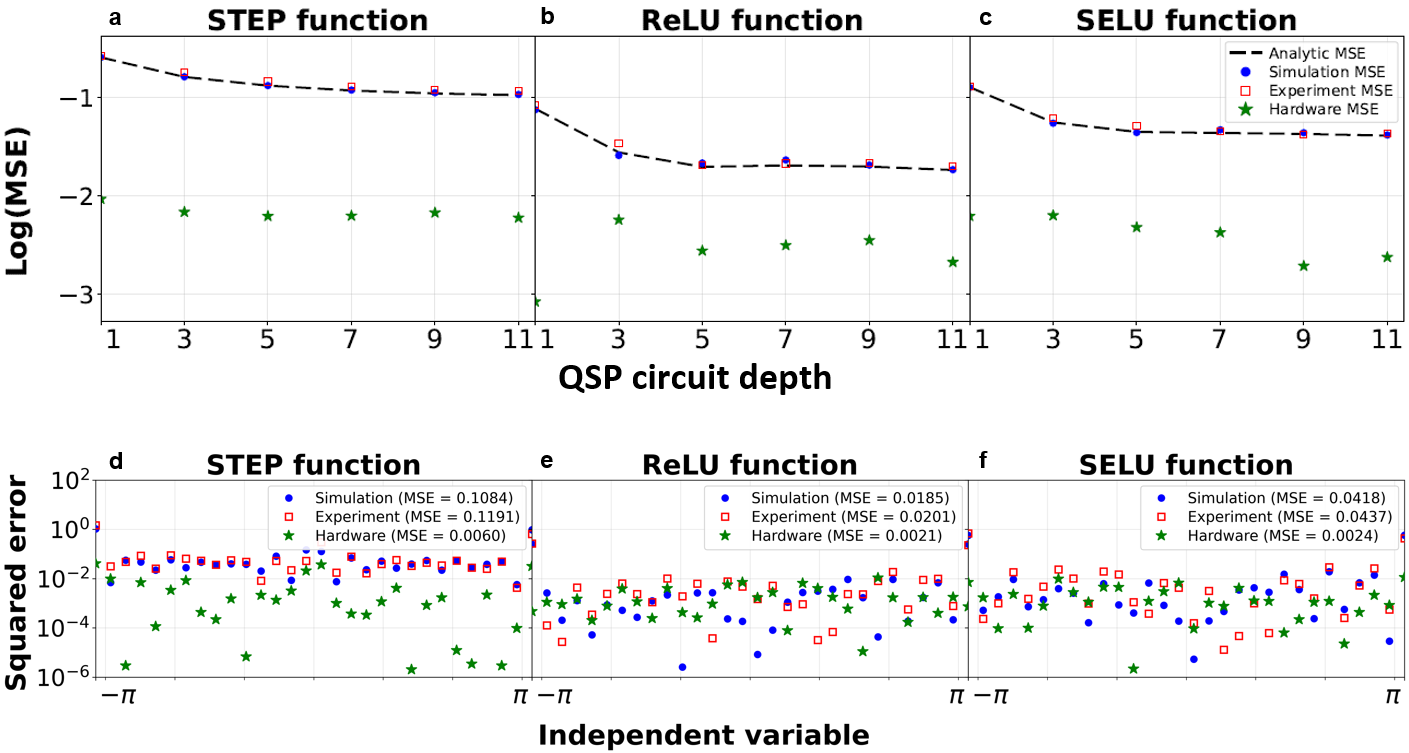}
  \caption{\label{fig:mse}
  \textbf{Error analysis of the photonic QSP implementation.}
  \textbf{a--c,} MSE on a logarithmic scale as a function of QSP circuit
  depth $L=1,3,5,7,9,11$ for STEP, ReLU and SELU, respectively.
  The black dashed lines show the analytic MSE obtained from the analytic
  prediction of the QSP using the characterized
  directional-coupler splitting ratios, blue circles show the numerically
  simulated MSE of the PIC, red squares show the experimental MSE and
  green stars show the hardware MSE, defined as the MSE between the
  experimental result and the analytic prediction of QSP at the same depth.
  Each experimental point is obtained from 30 input values, with
  $N_{\mathrm{shots}}=5000$ heralded single-photon detection events per
  input value.
  \textbf{d--f,} Squared error resolved across the 30 sampled input values
  at $L=11$ for STEP, ReLU and SELU, respectively. Blue circles show
  $|f_{\mathrm{sim}}(x)-f_{\mathrm{target}}(x)|^2$, red squares show
  $|f_{\mathrm{exp}}(x)-f_{\mathrm{target}}(x)|^2$ and green stars show
  the hardware squared error
  $|f_{\mathrm{exp}}(x)-f_{\mathrm{analytic}}(x)|^2$.
  The MSE values shown in the legends of panels d--f are the
  averages of the corresponding squared errors over the 30 sampled input
  values.}
\end{figure}

Figure~\ref{fig:mse}(a--c) shows the MSE as a function of $L$ for the
three target functions. For all three target functions, the analytic,
simulated, and experimental MSEs decrease with increasing $L$, reflecting
the improved polynomial expressivity of deeper QSP circuits. At the
maximum accessible depth $L=11$, the experimental MSEs relative to the
targets are $0.119$, $0.020$, and $0.044$ for STEP, ReLU, and SELU,
respectively, while the corresponding analytic values are $0.107$,
$0.018$, and $0.041$. The hardware-induced MSE, defined as the MSE between the experimental
result and the analytic prediction of QSP,
is $0.006$, $0.003$, and $0.003$ for STEP, ReLU, and SELU, respectively,
substantially below the corresponding errors relative to the target
functions. Thus, at the experimentally accessible depths, the dominant
contribution to the error relative to the target is the intrinsic
finite-depth QSP approximation rather than inaccurate execution of the
programmed transformation by the photonic processor. Increasing the
accessible QSP depth should therefore improve function approximation
accuracy as long as hardware errors remain below the finite-depth
approximation error. The approximation error depends strongly on the
target function and is largest for STEP, whose sharp transition requires
higher-order harmonics for accurate approximation. Meanwhile, the
hardware contribution remains small and relatively insensitive to the
programmed function.

The squared errors at each sampled input value for $L=11$ are shown in
Fig.~\ref{fig:mse}(d--f). For each input $x_n$, we evaluate
$|f_{\mathrm{exp}}(x_n)-f_{\mathrm{target}}(x_n)|^2$,
$|f_{\mathrm{sim}}(x_n)-f_{\mathrm{target}}(x_n)|^2$, and
$|f_{\mathrm{exp}}(x_n)-f_{\mathrm{analytic}}(x_n)|^2$ for the
experimental, simulated, and hardware errors, respectively. The
deviation of the experiment from the analytic prediction of the
QSP remains smaller than the deviation from the target over
most of the sampled domain, further showing that the dominant error at
this depth arises from the finite-depth function approximation rather
than the physical implementation.

Our PIC behavior contrasts with the error characteristics of the matter-qubit realization of QSP. In the trapped-ion realization~\cite{bu2025}, the experimental error is limited at shallow depths by
operational imperfections in the pulse sequence, and at depths beyond
$L\approx180$, the measured results diverge from the noiseless prediction, 
because time-dependent decoherence accumulates over the lengthening sequence and
corrupts the realized unitary. In contrast, on the photonic platform, the dominant error
mechanism is instead photon loss, which is heralded through the absence of
a detection event. Provided the loss is the same for both waveguide modes,
lost photons reduce the detected count rate but do not corrupt the
transformation applied to the detected photons. Residual path-dependent loss would bias the post-selected statistics, but the smallness of the hardware MSE in
Fig.~\ref{fig:mse} shows that this effect is minor at the accessible
depths. The achievable accuracy is therefore predominantly set by the
QSP approximation. A shallow-depth comparison shows that the photonic implementation has experimental
$\log_{10}\mathrm{MSE}$ values of $-0.92$, $-1.70$, and $-1.36$ for STEP,
ReLU, and SELU at $L=11$, respectively, while the trapped-ion implementation has experimental
$\log_{10}\mathrm{MSE}$ values of $\approx-1.0$ for all three functions at $L=15$. Meanwhile, the MSE for the STEP function is slightly larger in the photonic implementation at $L=11$ than in the trapped-ion implementation at $L=15$, consistent with the greater circuit depth required to approximate the sharp transition accurately.
This low hardware-induced error persists across all accessible circuit depths.
As shown in Fig.~\ref{fig:mse}, the hardware MSE remains in the range
$10^{-3}$ to $10^{-2}$ for all functions and circuit depths, lies well
below the corresponding total MSE, and does not grow with $L$.

More broadly, the key distinction lies in the \emph{type} of error rather
than its magnitude on any platform. Matter-qubit processors, such as
trapped-ion and superconducting architectures, are ultimately
decoherence-limited at large depth~\cite{bu2025}, and decoherence degrades the
fidelity of the realized operation in a way that cannot be recovered by post-selection in the absence of active error correction. Photon loss, by
contrast, is heralded through the absence of a detection event, and therefore imposes primarily a count-rate cost rather than a fidelity cost, and can be
mitigated by brighter sources and lower-loss components.

\section{Discussion}

The $SU(2)$ correspondence, detailed in Methods and Supplementary
Section~1, establishes that the native two-mode operations of programmable interferometric PICs realize the mathematical structure required by QSP, enabling QSP sequences to be mapped directly onto the native operations of the photonic processor. This mapping provides an architectural link between QSP algorithms and programmable photonic hardware. Because QSP underlies QSVT and a broad class of
quantum algorithms~\cite{Gilyen2019,Martyn2021,dong2022}, this implementation is a first step toward photonic realizations of QSP-enabled quantum algorithms. The demonstrated mapping also shows that
programmable nonlinear functions of encoded variables can be generated
algorithmically on linear photonic hardware without a
physical optical nonlinearity.

The resulting functionality is not limited to the STEP, ReLU, and SELU functions demonstrated here. More broadly, the same QSP framework can synthesize other polynomial or trigonometric transformations that form computational primitives for tasks including phase processing, filtering, Hamiltonian simulation, and matrix-function transformations~\cite{Gilyen2019,Martyn2021,wang2023qpp,dong2022,Martyn2023,bu2025}. In the present implementation, an external scalar variable can be mapped to the phase-encoding domain and encoded as the input $x$, after which the programmed QSP sequence produces the measured response $f(x)=P(0)-P(1)$. Beyond these quantum-algorithmic applications, the demonstrated nonlinear mapping could also serve as a programmable nonlinear stage within larger photonic information-processing architectures. For example, photonic neural networks typically alternate linear optical transformations with nonlinear activation functions~\cite{Shen2017,Williamson2020}. The output of a preceding photonic layer could be converted to a scalar variable, normalized to the QSP input domain, and encoded as the phase variable $x$. The QSP circuit could then implement a prescribed nonlinear function of this variable, with the output obtained from $f(x)=P(0)-P(1)$. In this setting, the nonlinear function is determined by the programmed QSP phase sequence rather than by a fixed material or device response. The alternating signal-encoding and programmable-rotation structure is also closely related to single-qubit data-reuploading quantum neural networks, with QSP providing analytically synthesized rather than trained phase parameters~\cite{perezsalinas2020,Ono2023,Mauser2026,Cerezo2021}. 

Photonic and matter-qubit QSP implementations face different scaling
limitations. In the present photonic platform, increasing circuit depth is
primarily constrained by photon loss, whereas matter-qubit
implementations ultimately face accumulated operational errors and
decoherence~\cite{bu2025}. Future photonic QSP implementations may
benefit from high-speed electro-optic phase control based on thin-film
lithium niobate, which combines low optical loss, low switching voltage,
and gigahertz-scale bandwidth~\cite{Hu2025TFLNReview}.

Although the present demonstration is limited to single-qubit QSP and
the circuit depth supported by the available processor, larger
programmable photonic systems could support deeper and more complex QSP
circuits. Recent
demonstrations of programmable photonic quantum processors and
loss-tolerant photonic architectures show continuing
progress in scaling photonic quantum information
processing~\cite{Madsen2022,Bartolucci2023}.
Larger meshes, higher-dimensional path encodings and multi-qubit
extensions could enable photonic implementations of QSP-
and QSVT-enabled quantum algorithms~\cite{Martyn2021,Aharonovich2026}.
\section{Conclusion}

We experimentally demonstrated a hardware-native implementation of trigonometric QSP on a programmable PIC and used it to synthesize nonlinear functions on a linear photonic processor.
Using a dual-rail single-photon qubit encoded in a reconfigurable linear optical mesh, we experimentally synthesized STEP, ReLU, and SELU functions through analytically generated QSP phase sequences, with measurements closely matching the corresponding theoretical predictions across all experimentally accessible circuit depths.

Our results further show that programmable integrated photonics offers a distinct operating regime for QSP. At our experimentally accessible circuit depths, the implemented transformations achieve high accuracy with hardware-induced errors substantially smaller than the intrinsic approximation error. At larger depths, photon loss is expected to become an increasingly important limitation and represents a key challenge for scaling photonic QSP. However, because loss events are heralded through the absence of a successful detection event, they primarily reduce the successful sampling rate rather than directly producing erroneous measurement outcomes. As photonic hardware continues to advance through lower-loss components, brighter single-photon sources, more efficient detectors, and larger programmable interferometric meshes, substantially deeper QSP implementations are expected to become experimentally accessible~\cite{Senellart2017,Uppu2021,Tomm2021,EsmaeilZadeh2021}.

 These results establish programmable PICs as a hardware-native platform
for QSP and provide a foundation for larger-scale photonic implementations
of QSP-enabled quantum algorithms and programmable nonlinear transformations.
\section*{Methods}

\subsection*{Experimental platform}

The experiments were performed on the Quandela Belenos programmable
photonic processor, a 24-mode programmable linear-optical processor
comprising a universal interferometric mesh. The photonic circuit is
fabricated in an alumino-borosilicate glass substrate and occupies a
footprint of $15\times134~\mathrm{mm}^2$~\cite{Barzaghi2025}.
The platform architecture,
combining a quantum-dot single-photon source with a reconfigurable
universal interferometer, is described in Ref.~\cite{Maring2024}. The QSP circuit was
implemented using two path-encoded waveguide modes, while the remaining
modes were left inactive. Single photons generated by the integrated
quantum-dot source~\cite{Maring2024} were injected into the upper waveguide of the
dual-rail circuit, and the output probabilities were reconstructed from
repeated single-photon measurements at the two logical output modes.
According to the platform specifications, the source exhibited a
second-order correlation of $g^{(2)}(0)=0.013$, a photon
indistinguishability of $0.902$, and an overall system transmittance of
$0.058$. On-chip phases are set electrically via thermo-optic phase shifters, with
the conversion from programmed phases to drive voltages handled by
Quandela's transpilation~\cite{fyrillas2024clearbox}.

\subsection*{$SU(2)$ mapping of QSP to the photonic circuit}
\label{sec:su2_mapping}

QSP and its extensions have been developed and
reviewed extensively in the literature~\cite{Martyn2021,motlagh2024,joven2026scalable}.
In particular, the sequence of
general $SU(2)$ signal-processing rotations employed here can be viewed
as a single-qubit instance of generalized QSP~\cite{motlagh2024}.

The QSP circuit of Eq.~(\ref{eq:qsp}) is implemented by mapping each
single-qubit operation onto a two-mode linear optical transformation in
the dual-rail encoding, where the logical basis states correspond to a
single photon occupying one of two spatial modes.
$|0\rangle_L\equiv|1,0\rangle$ denotes the photon in the upper
waveguide, and $|1\rangle_L\equiv|0,1\rangle$ the photon in the lower
waveguide, where $|n_0,n_1\rangle$ labels the photon-number state of the
upper and lower modes, respectively. In this subspace, a $2\times2$
unitary acting on the two optical modes implements the corresponding
single-qubit unitary on the logical basis, so it suffices to show that
the mode transformations available on the processor generate the
operators appearing in Eq.~(\ref{eq:qsp}).

In the single-photon dual-rail subspace, lossless two-mode
interferometry provides the spin-$1/2$ representation of
$SU(2)$~\cite{Campos1989}, with the Schwinger generators $\hat J_k$
reducing to $\sigma_k/2$ for $k=x,y,z$. Consequently, optical phase
differences between the two modes implement the $R_z$ rotations
required by QSP, and the two-mode interferometric mixing implements
$R_y$ rotations.

\textit{Two-mode interferometry as a native architecture for QSP.}
The connection between two-mode interferometry and QSP can be seen
directly from the beam-splitter transformation. Consider a general
lossless beam splitter
\begin{equation}
\mathrm{BS}(\chi)
=
\exp\left[
-\frac{i}{2}
\left(
\chi^{*}\hat a\hat b^{\dagger}
+
\chi\hat a^{\dagger}\hat b
\right)
\right],
\qquad
\chi=\beta e^{i\alpha},
\label{eq:bs_general}
\end{equation}
where $\hat a$ and $\hat b$ are the bosonic annihilation operators of
the two optical modes, $\beta$ is the mixing angle, and $\alpha$
determines the phase of the mode coupling. In the single-photon
dual-rail subspace, this transformation corresponds to an $SU(2)$
rotation whose axis in the equatorial plane is determined by
$\alpha$. As derived in the Supplementary Information, a sequence of
$L+1$ such two-mode interferometric transformations with programmable
phases can be rearranged into the standard QSP structure. Its action on
the mode operators takes the form
\begin{align}
\begin{pmatrix}
\hat a_{\beta,\boldsymbol{\alpha}}\\
\hat b_{\beta,\boldsymbol{\alpha}}
\end{pmatrix}
&=
R_z(\Phi_0)
\prod_{j=1}^{L}
R_x(\beta)R_z(\Phi_j)
\begin{pmatrix}
\hat a\\
\hat b
\end{pmatrix}
\nonumber\\[3pt]
&=
\begin{pmatrix}
P(s) &
-i\sqrt{1-s^2}\,Q(s)
\\
-i\sqrt{1-s^2}\,Q^{*}(s) &
P^{*}(s)
\end{pmatrix}
\begin{pmatrix}
\hat a\\
\hat b
\end{pmatrix},
\qquad
s=\cos(\beta/2),
\label{eq:bs_qsp}
\end{align}
where $P(s)$ is a polynomial of degree at most $L$ and $Q(s)$ is a
polynomial of degree at most $L-1$. The phases $\{\Phi_j\}$ are
determined by the beam-splitter phases $\{\alpha_k\}$. This is the
characteristic polynomial structure of QSP and shows directly that
programmable two-mode interferometry provides a native architecture
for realizing QSP.

The general construction above assumes beam-splitter transformations
with a controllable mixing angle $\beta$ and coupling phase $\alpha$.
On the Belenos processor, the directional couplers are instead fixed at
a nominal 50:50 splitting ratio, corresponding to $\beta=\pi/2$ in
Eq.~(\ref{eq:bs_general}). The required programmable two-mode
rotations are therefore synthesized using MZIs composed of two fixed
directional couplers and programmable phase shifters. The beam-splitter
derivation establishes the underlying correspondence between two-mode
interferometry and QSP, while the MZI construction provides the
hardware realization of the required rotations on the present PIC.
The specific MZI realization used on Belenos, including the effect of
the fabricated directional-coupler splitting ratios and phase
compilation, is described below and in the Supplementary Information.

\textit{Diagonal rotations.} Phase shifters applying phases
$\varphi_0$ and $\varphi_1$ to the two modes implement
\begin{equation}
\mathrm{diag}(e^{i\varphi_0},e^{i\varphi_1})
=
e^{i(\varphi_0+\varphi_1)/2}
R_z(\varphi_1-\varphi_0),
\end{equation}
so every $R_z$ rotation, including both the signal-processing phases
$R_z(\phi_j)$ and the signal unitary $R_z(x)$, is realized exactly by a
phase difference between the two waveguides, up to a global phase. In
the experiment, each such rotation is programmed using a single
physical phase shifter on the upper mode with the corresponding phase
difference.

\textit{$R_y$ rotations.} Each $R_y(\theta_j)$ rotation is implemented
by a programmable MZI comprising two directional couplers together with
programmable internal and output phase shifters. The complete MZI
transfer-matrix derivation is provided in the Supplementary Information.

\textit{Fabricated couplers.} On the Belenos processor, the directional
couplers deviate from the ideal 50:50 splitting ratio. The phase settings
of each physical MZI are therefore determined using the
characterized splitting ratios of the fabricated directional
couplers. Details of this decomposition are provided in the
Supplementary Information.

\textit{Phase aggregation.} The phase contributions associated with
adjacent $R_z(\phi_j)$ and $R_z(x)$ rotations are combined analytically
to obtain the equivalent differential phase programmed between
neighboring MZIs. Details of the phase aggregation are provided in the
Supplementary Information.

Because the individual photonic transformations reproduce the
corresponding $R_y$ and $R_z$ operations, their composition implements
the complete QSP sequence,
\begin{equation}
U_{\mathrm{PIC}}(x)
=
e^{i\gamma(x)}
A(\theta_0,\phi_0)
\prod_{j=1}^{L}
R_z(x)A(\theta_j,\phi_j)
=
e^{i\gamma(x)}W(x),
\label{eq:pic_qsp_equivalence}
\end{equation}
where $\gamma(x)$ is an overall optical phase and therefore does not
affect the measured probabilities. Thus the QSP phases $\phi_j$ and the signal-encoding phase $x$ in
Eq.~(\ref{eq:qsp}) are mapped directly to the programmable optical
phase differences of the PIC. These phases are set experimentally by
electrically controlling the on-chip thermo-optic phase shifters. The corresponding Heisenberg-picture
derivation for the complete sequence and the mapping of the output
photon-number difference to the QSP Pauli-$Z$ measurement are provided
in the Supplementary Information.

Before execution on the QPU, the programmed mesh unitary was numerically
verified against the corresponding target QSP transformation for every
input value $x$ using the Perceval software platform~\cite{Heurtel2023}. Details of the circuit compilation and numerical simulations
are provided in the Supplementary Information.

\subsection*{QSP phase synthesis}

For each target function and QSP circuit depth $L$, the target function
is first approximated by a finite trigonometric polynomial. The
corresponding QSP phase sequence
$\{(\theta_j,\phi_j)\}_{j=0}^{L}$ is then computed classically using the
QSP angle-finding module provided by Paddle
Quantum~\cite{paddlequantum2023}. The angle-finding algorithm takes the
target polynomial as input and returns the phase sequence implementing
the desired polynomial transformation. The resulting angles are mapped
directly to the programmable phase shifters implementing the
signal-processing unitaries $A(\theta_j,\phi_j)$ on the photonic
processor.
The target-function definitions and the complete QSP phase sequences
used for STEP, ReLU, and SELU at $L=3$, $7$, and $11$ are provided in
the Supplementary Information.
\section*{Acknowledgments}
This work is supported by the U.S. Department of Energy, Office of Science, Advanced Scientific Computing Research, under contract number DE-SC0025384 and the National Science Foundation under Awards ExpandQISE-2329027 and ECCS-2533541.


\bibliographystyle{unsrt}
\bibliography{refs}

\end{document}